\pdfoutput=1
\documentclass[11pt,twoside,a4paper,cmspaper,final,collab]{cms-tdr}
\def\svnVersion{000206f}\def\svnDate{2025/04/02}\def\cmsCernNoTag{CERN-EP-2024-176}\def\cmsCernDate{\today}\def\cmsMessage{Published in the Journal of High Energy Physics as \href{http://dx.doi.org/10.1007/JHEP02(2025)195}{\doi{10.1007/JHEP02(2025)195}.}}
\begin{document}\cmsNoteHeader{HIN-21-014}

\hyphenation{had-ron-i-za-tion}
\hyphenation{cal-or-i-me-ter}
\hyphenation{de-vices}

\newcommand{\sqrts}{ \ensuremath{\sqrt{s}}\xspace}
\newcommand{\RAA}{\ensuremath{R_{\mathrm{AA}}}\xspace}
\newcommand{\TAA}{\ensuremath{T_{\mathrm{AA}}}\xspace}
\newcommand{\PbPb}{\ensuremath{\mathrm{PbPb}}\xspace}
\newcommand{\pp}{\Pp{}\Pp}
\newcommand{\pPb}{\Pp{}\ensuremath{\mathrm{Pb}}\xspace}
\newlength\cmsTabSkip\setlength{\cmsTabSkip}{1ex}
\newcommand{\Pphi}{\PGfP{1020}}
\newcommand{\Bzerosdecayshort}{\ensuremath{\PBzs \to \PJGy (\Pgmp\Pgmm)\,\Pphi (\PKp \PKm)}\xspace}
\newcommand{\Bplusdecayshort}{\ensuremath{\PBp \to \PJGy (\Pgmp\Pgmm)\,\PKp}\xspace}
\newcommand{\Bzerosdecay}{\ensuremath{\PBzs \to \PJGy\,\Pphi}\xspace}
\newcommand{\Bplusdecay}{\ensuremath{\PBp \to \PJGy\,\PKp}\xspace}
\newcommand{\Jpsidecay}{\ensuremath{\PJGy\to\PGmp\PGmm}\xspace}
\newcommand{\dstardecay}{\ensuremath{\PDst \to \PDz (\PK\PGp)\,\PGp\xspace}}
\newcommand{\dstardecayfive}{\ensuremath{\PDst \to \PDz (\PK\PGp\PGp\PGp)\,\PGp\xspace}}
\newcommand{\phidecay}{\ensuremath{\Pphi\to \PKp \PKm \xspace}}
\newcommand{\acceff}{\ensuremath{\alpha\,\epsilon}\xspace}
\newcommand{\acceffinv}{\ensuremath{1/(\acceff)}\xspace}
\newcommand{\meanacceffinv}{\ensuremath{\langle\acceffinv\rangle}\xspace}
\newlength\cmsFigWidth
\newcommand{\sPlot}{\hbox{$_s$}\mathcal{P}lot}
\setlength\cmsFigWidth{0.65\textwidth}
\providecommand{\cmsLeft}{left\xspace}
\providecommand{\cmsRight}{right\xspace}

\cmsNoteHeader{HIN-21-014}
\title{Bottom quark energy loss and hadronization with \texorpdfstring{\PBp}{B+} and \texorpdfstring{\PBzs}{B0s} nuclear modification factors using \texorpdfstring{\pp}{pp} and \texorpdfstring{\PbPb}{PbPb} collisions at \texorpdfstring{$\sqrtsNN=5.02\TeV$}{sqrtsNN=5.02 TeV}
}

\date{\today}
\abstract{The production cross sections of \PBzs and \PBp mesons are reported in proton-proton (\pp) collisions recorded by the CMS experiment at the CERN LHC with a center-of-mass energy of 5.02\TeV. The data sample corresponds to an integrated luminosity of 302\pbinv. The cross sections are based on measurements of the \Bzerosdecayshort and \Bplusdecayshort decay channels. Results are presented in the transverse momentum (\pt) range 7--50\GeVc and the rapidity interval $\abs{y} < 2.4$ for the \PB mesons. The measured \pt-differential cross sections of \PBp and \PBzs in \pp collisions are well described by fixed-order plus next-to-leading logarithm perturbative quantum chromodynamics calculations. Using previous \PbPb collision measurements at the same nucleon-nucleon center-of-mass energy, the nuclear modification factors, \RAA, of the \PB mesons are determined. For $\pt > 10\GeVc$, both mesons are found to be suppressed in \PbPb collisions (with \RAA values significantly below unity), with less suppression observed for the \PBzs mesons. In this \pt range, the \RAA values for the \PBp mesons are consistent with those for inclusive charged hadrons and \PDz mesons. Below $10\GeVc$, both \PBp and \PBzs are found to be less suppressed than either inclusive charged hadrons or \PDz mesons, with the \PBzs \RAA value consistent with unity. The \RAA values found for the \PBp and \PBzs are compared to theoretical calculations, providing constraints on the mechanism of bottom quark energy loss and hadronization in the quark-gluon plasma, the hot and dense matter created in ultrarelativistic heavy ion collisions.}

\hypersetup{%
pdfauthor={CMS Collaboration},%
pdftitle={Bottom quark energy loss and hadronization with B+ and B0s nuclear modification factors in PbPb collisions at 5.02 TeV with CMS},%
pdfsubject={CMS},%
pdfkeywords={CMS, quark-gluon plasma, bottom physics}}

\maketitle
\section{Introduction} 
Ultrarelativistic collisions of heavy ions create extreme temperatures and energy densities.
As predicted by lattice quantum chromodynamics (QCD) calculations~\cite{Karsch:2003jg},
such conditions produce a state of matter referred to as the quark-gluon plasma (QGP)~\cite{QGP1, QGP2}, in which the relevant degrees of freedom of the system can be described by the interactions of quarks and gluons~\cite{Busza:2018rrf}.
Heavy quarks can be used as probes of the
QGP~\cite{Andronic:2015wma, Dong:2019byy, Apolinario:2022vzg}. The study of their interactions with the medium they traverse, which include both elastic collisions and medium-induced
gluon radiation~\cite{Eloss1, Baier:2000mf, Chatrchyan:2011sx, Aad:2010bu}, provide insights into the transport properties of the QGP. 
While data relevant to the energy loss of heavy flavor quarks
are available from RHIC and LHC experiments~\cite{Andronic:2015wma},
comprehensive measurements sensitive to the influence of the QGP medium on the transport of
flavor-identified heavy hadrons are crucial for a deeper understanding of the hadronization mechanism. Such measurements are needed to extract transport properties, including, e.g., the heavy quark diffusion constant~\cite{Dong:2019byy,Apolinario:2022vzg} and the jet quenching parameter~\cite{Apolinario:2022vzg}.

For bottom quarks, in particular, there is significant theoretical uncertainty associated with 
how the presence of the QGP medium affects the hadronization process~\cite{Rapp:2018qla}. The QGP is known to have an abundant strangeness content~\cite{Abelev:2013xaa, Agakishiev:2011ar, Abelev:2008zk, Arsene:2009jg, Adamczyk:2017wsl, Adare:2015ema, ALICE:2017jyt} since its temperature is above the strange quark
mass~\cite{PhysRevLett.48.1066}. This abundance may enhance the production of \PBzs mesons in the QGP medium, relative to the non-strange \PBp mesons, through a quark recombination process~\cite{Molnar:2003ff, Greco:2003mm, Greco:2003vf, He:2014cla}.
Studies of open-charm~\cite{Acharya:2018hre} and
open-bottom~\cite{BsPbPbCMS} mesons in lead-lead (\PbPb) collisions at a nucleon-nucleon (NN) center-of-mass energy of $\sqrtsNN=5.02\TeV$ at the LHC hint at an enhanced production of strange-charm and strange-bottom hadrons. Recently, this line of studies has been extended to 
heavier \PBc~\cite{CMS:2022sxl} and exotic $\PX(3872)$~\cite{CMS:2021znk} 
hadron production in \PbPb collisions.
Reference~\cite{CMS:2024krd} presents a summary of previous CMS work on how the QGP medium affects particle production at the LHC, as well as a more general overview of the CMS heavy ion program.

Measurements comparing various flavors of bottom hadrons have the potential to yield new insights regarding the hadronization of heavy quarks.
Such measurements are challenging because of the relatively small production cross sections, large and varied background sources, and detector resolution
limitations. In this paper, we present a study of the production rates of the \PBp and \PBzs mesons in proton-proton (\pp) collisions at $\sqrts=5.02\TeV$,
using a data set corresponding to $302 \pm 6\pbinv$~\cite{CMS-PAS-LUM-19-001} recorded with the CMS detector at the LHC in 2017.
The \PB mesons are reconstructed via the exclusive decay channels \Bzerosdecay and \Bplusdecay, with \Jpsidecay and \phidecay.
The \pp cross sections are presented as
functions of the transverse momentum (\pt) of the meson measured in the rapidity interval
$\abs{y}<2.4$ and are compared to fixed-order plus next-to-leading logarithm (FONLL)
calculations~\cite{FONLLcharmbottomPP1, FONLLcharmbottomPP2, FONLLcharmbottomPP3}.
The new results are complementary to \PB meson measurements in \pp collisions at $\sqrts=5.02$, 7, 8, and
13\TeV~\cite{BpPbPbCMS, BsPbPbCMS, CMSBmesonpp,Chatrchyan:2011pw,Chatrchyan:2011vh,ATLAS:2013cia,LHCb:2012sng,LHCb:2013JHEP,LHCb:2011leg,Aaij:2012dd,Khachatryan:2014nfa,Aaij:2015fea,Aaij:2014ija,ATLAS:2019jpi,Khachatryan:2016csy,LHCb:2017vec,LHCb:2019tea},
in proton-lead collisions at $\sqrtsNN=5.02$ and
8.16\TeV~\cite{Khachatryan:2015uja,LHCb:2019avm}, and in \PbPb collisions at
$\sqrtsNN=5.02\TeV$~\cite{BsPbPbCMS,BpPbPbCMS,CMS:2021mzx}.
The published \PbPb results~\cite{CMS:2021mzx} for \PBp and \PBzs mesons are used to calculate the respective nuclear modification factors $\RAA$, \ie, the meson yield ratio in nucleus-nucleus (AA) and \pp interactions normalized by the number of inelastic NN collisions via the relation
\begin{equation}
 \RAA (\pt) = \frac{1}{\TAA} \frac{{\rd}N^{\PBp,\PBzs}_\PbPb}{{\rd}\pt} \bigg/ \frac{{\rd}\sigma^{\PBp,\PBzs}_{\pp}}{{\rd}\pt},
 \label{eq:RAA}
\end{equation}
where \TAA is the average number of NN binary collisions per \PbPb interaction divided by the NN total
inelastic cross section, which can be interpreted as the NN-equivalent integrated luminosity per heavy ion collision~\cite{CMS:2021mzx}.
The reported \RAA results are for collisions with 0--90\% centrality, corresponding to
the 90\% of collisions having the largest overlap of the two nuclei~\cite{Chatrchyan:2011sx,CMS:2021mzx}.
Throughout this paper, unless otherwise specified, the $y$ and \pt variables refer to the \PB mesons.
All references to \PB mesons in the text also include the corresponding charge-conjugate
state, although the quoted cross sections correspond to individual states \ie, the
average of the two yields), as was done for the
previous \PbPb analysis~\cite{CMS:2021mzx}.
Tabulated results are provided in the HEPData record for this analysis~\cite{hepdata}.

\section{The CMS apparatus}
The central feature of the CMS apparatus is a superconducting solenoid of 6\unit{m} internal diameter, providing a magnetic field of 3.8\unit{T}. Within the solenoid volume are a silicon pixel and strip tracker, a lead tungstate crystal electromagnetic calorimeter, and a brass and scintillator hadron calorimeter, each composed of a barrel and two endcap sections.
The silicon tracker measures charged particles within the pseudorapidity range $\abs{\eta} < 3.0$. During the LHC running period when the data used in this paper were recorded, the silicon tracker consisted of 1856 silicon pixel and 15\,148 silicon strip detector modules. Details on the pixel detector can be found in Ref.~\cite{Phase1Pixel}. For nonisolated particles of $1 < \pt < 10\GeVc$ and $\abs{\eta} < 3.0$, the track resolutions are typically 1.5\% in \pt and 20--75\mum in the transverse impact parameter~\cite{DP-2020-049}.
Muons are measured in the range $\abs{\eta} < 2.4$, with detection planes made using three
technologies: drift tubes, cathode strip chambers, and resistive plate chambers. The
single-muon trigger efficiency exceeds 90\% over the full $\eta$ range, and the efficiency
to reconstruct and identify muons is greater than 96\%. Matching muons to tracks measured
in the silicon tracker results in a relative \pt resolution, for muons with $\pt <
100\GeVc$, of 1\% in the barrel and 3\% in the endcaps. The \pt resolution in the barrel
is better than 7\% for muons with $\pt < 1\TeVc$~\cite{CMS:2018rym}. The hadron forward (HF) calorimeters, situated 11.2\unit{m} from the interaction region on both sides, with the coverage in the range $3.0 < \abs{\eta} < 5.2$, utilize steel for absorption and quartz fibers as sensitive material. Information from the HF is used for determining the centrality.

Events of interest are selected using a two-tiered trigger system. The first level, composed of custom hardware processors, uses information from the calorimeters and muon detectors to select events at a rate of around 100\unit{kHz} within a fixed latency of 4\mus~\cite{CMS:2020cmk}. The second level, known as the high-level trigger, consists of a farm of processors running a version of the full event reconstruction software optimized for fast processing, and reduces the event rate to around 1\unit{kHz} before data storage~\cite{CMS:2016ngn}. A detailed description of the CMS experiment and
coordinate system can be found in Ref.~\cite{CMS:2008xjf}.

\section{Analysis method}
\subsection{Candidate reconstruction and selection}
\label{sec:selection}
The \pp collision events are collected with a first-level trigger requiring the presence
of two muon candidates, with no explicit momentum threshold. A subsequent high-level trigger required one of the muon candidates to be reconstructed using information from both the muon detectors and the inner tracker, while only information from the muon detectors is required for the other muon candidate. The dimuon candidates are further required to have an invariant mass within $1<m_{\Pgm\Pgm}<5\GeVcc$.

For the offline analysis, events must pass a set of selection criteria designed to
reject non-collision events and to select only inelastic hadronic
collisions~\cite{CMS:2016xef,Sirunyan:2017qhf}. Events are required to have at
least one reconstructed primary interaction vertex, formed by two or more tracks, with a distance from the center of the nominal interaction region of less than 25\cm along the beam axis. The shapes of the clusters in the pixel detector must be compatible with those expected from particles produced in a \pp collision~\cite{Khachatryan:2010xs}. To select inelastic hadronic collisions, the \pp events are also required to have at least one tower in both of the HF detectors with an energy deposit of more than $4\GeV$.

The signal is extracted following the same procedure as for the \PbPb results~\cite{CMS:2021mzx}, as will be described.
The same kinematic constraints are applied for the \pp and \PbPb samples to select muon candidates.
The muons are required to match the trigger-level muon candidates and to pass the \emph{soft muon} selection criteria~\cite{Chatrchyan:2012xi}.
Two muons of opposite charge, with an invariant mass within $\pm150\MeVcc$ of the world-average \PJGy meson mass~\cite{10.1093/ptep/ptac097}, are selected to reconstruct a \PJGy candidate.
All remaining tracks are considered kaon candidates with no attempt at particle ID.
The \Pphi meson candidates are formed with a common-vertex constraint between two oppositely charged tracks with $\pt > 500\MeVc$.
The resulting invariant mass, assuming the world-average charged-kaon mass~\cite{10.1093/ptep/ptac097} for each of the two tracks
is required to lie within $\pm15\MeVcc$ of the world-average \Pphi meson mass~\cite{10.1093/ptep/ptac097}.
Muon pairs (from \PJGy meson decays) and \Pphi meson candidates are combined to form \PBzs
meson candidates, requiring that they originate from a common vertex. Similarly, muon
pairs are combined with a single charged particle, assuming the \PKp mass, to form \PBp candidates.
Negative tracks are also included to account for the charge-conjugate state.
The \PB meson candidates are reconstructed within a fiducial region of
$1.5 < \abs{y} < 2.4$ for $\pt < 10\GeVc$ and $\abs{y} < 2.4$ for $\pt > 10\GeVc$,
which are the same regions as used in the \PbPb analysis~\cite{CMS:2021mzx}.
Muons from \PB decays at midrapidity have a reduced detector acceptance at low \pt,
therefore fiducial regions have been chosen in a \pt-dependent way.

Several samples of Monte Carlo (MC) simulated events are used to evaluate background components, signal efficiencies, and detector acceptance corrections. 
The simulations include signal samples containing only the \PB meson decay channels being measured and background samples of other \PQb hadron decay chains also involving \PJGy mesons.
These signal and background simulated samples in \pp collisions are generated with \PYTHIA{}8 v230~\cite{Sjostrand:2014zea}, 
with underlying event tunes
CP5~\cite{CMS:2019csb} (signal) and CUETP8M1~\cite{Khachatryan:2015pea} (background), respectively.
The resulting particles are propagated through a model of the CMS detector using the \GEANTfour package~\cite{Agostinelli:2002hh}. The decay of the \PB mesons is modeled with \EVTGEN 1.6.0~\cite{Lange:2001uf}, and the final-state photon radiation is simulated with \PHOTOS 2.0~\cite{Barberio:1990ms}. 
Pileup interactions (the number of concurrent interactions in the same bunch crossing)
are also simulated according to a mean pileup of 3.5 in the data set.

The final \PB candidate selection is determined via multivariate discriminators based on a
boosted decision tree (BDT)~\cite{Hocker:2007ht}. The same discriminating variables
are employed as in the \PbPb data analysis. These include the $\chi^{2}$ probability of the decay vertex (the probability for the muon tracks from the \PJGy meson decay and the other charged-hadron track(s) to originate from a common vertex),
the distance from the primary to the \PB decay vertices (normalized by its uncertainty), the pointing angle (the angle between the line segment connecting the primary and decay vertices and the momentum vector of the \PB meson), 
the \pt of the daughter charged-hadron track(s), and the transverse and longitudinal track
distances of closest approach to the primary vertex, normalized by their uncertainties.
For the \PBzs meson, an additional selection variable is used: the absolute difference between the reconstructed \Pphi meson invariant mass and its nominal value.
The BDT training is performed by employing the MC-simulated \PB signal samples, 
and background samples taken from data sidebands constructed using candidates with the invariant mass away from the \PB meson nominal mass, $0.25 < \abs{m_{\PBp}-m_{\PBp}^\mathrm{PDG}} < 0.30\GeVcc$ and $0.20 < \abs{m_{\PBzs}-m_{\PBzs}^\mathrm{PDG}} < 0.30\GeVcc$, where
$m_{\PBp}^\mathrm{PDG} = 5279.32\,(\pm0.14)\MeVcc$ and $m_{\PBzs}^\mathrm{PDG} = 5366.89\,(\pm0.19)\MeVcc$ are the world-average masses given by the Particle Data Group (PDG)~\cite{10.1093/ptep/ptac097}. The signal samples are scaled to the expected number of \PB candidates calculated by the measured integrated luminosity and cross sections predicted by FONLL perturbative QCD calculations~\cite{FONLLcharmbottomPP3}.
The BDT selection is individually optimized for each meson and \pt range to maximize the figure of merit $S/\sqrt{S+B}$.  Here, $S$ and $B$ are the number of signal and background candidates, respectively, in the signal region after applying the selection criteria. The signal region is defined as $|m_{\PB}-m_{\PB}^\mathrm{PDG}| < 0.08\GeVcc$.

\subsection{Differential cross sections}
The raw \PB meson signal yields are extracted from data using an extended unbinned maximum
likelihood fit to the invariant mass spectra, following the same procedure and fit
functions as for the \PbPb analysis~\cite{CMS:2021mzx}. The signal shape is modeled by a
sum of two Gaussian functions, with parameters determined from the MC simulation, except for the common mean and the overall signal yield, which are free parameters of the fit.
The fit also includes an additional free scaling parameter to account for a potential resolution difference between data and simulation.
The combinatorial background, from uncorrelated combinations of \PJGy candidates with
other particles, gives rise to a falling contribution in the invariant mass spectrum that
is modeled by an exponential function. 

\begin{figure}[!ht]
\centering
\includegraphics[width=.49\textwidth]{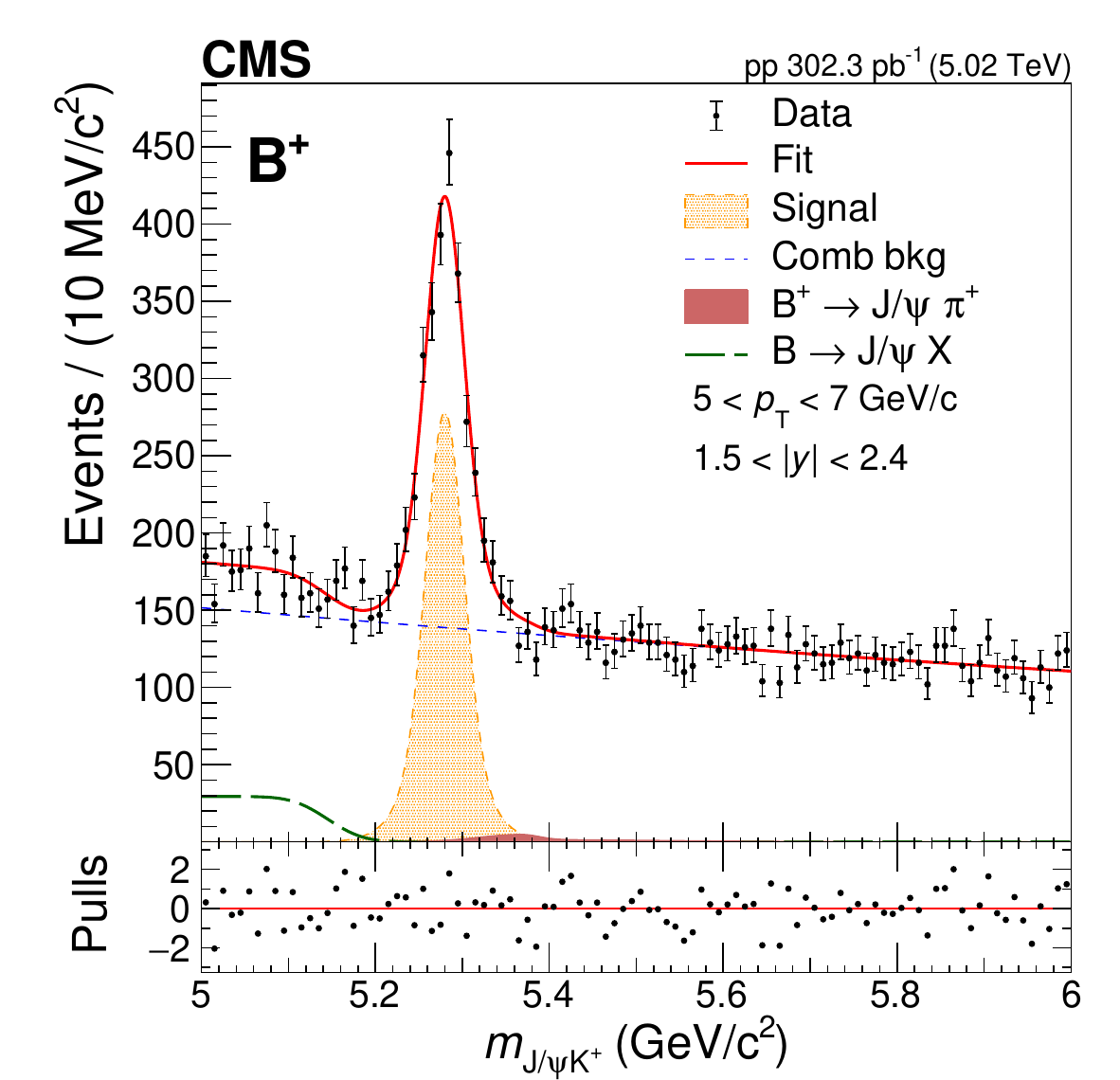}
\includegraphics[width=.49\textwidth]{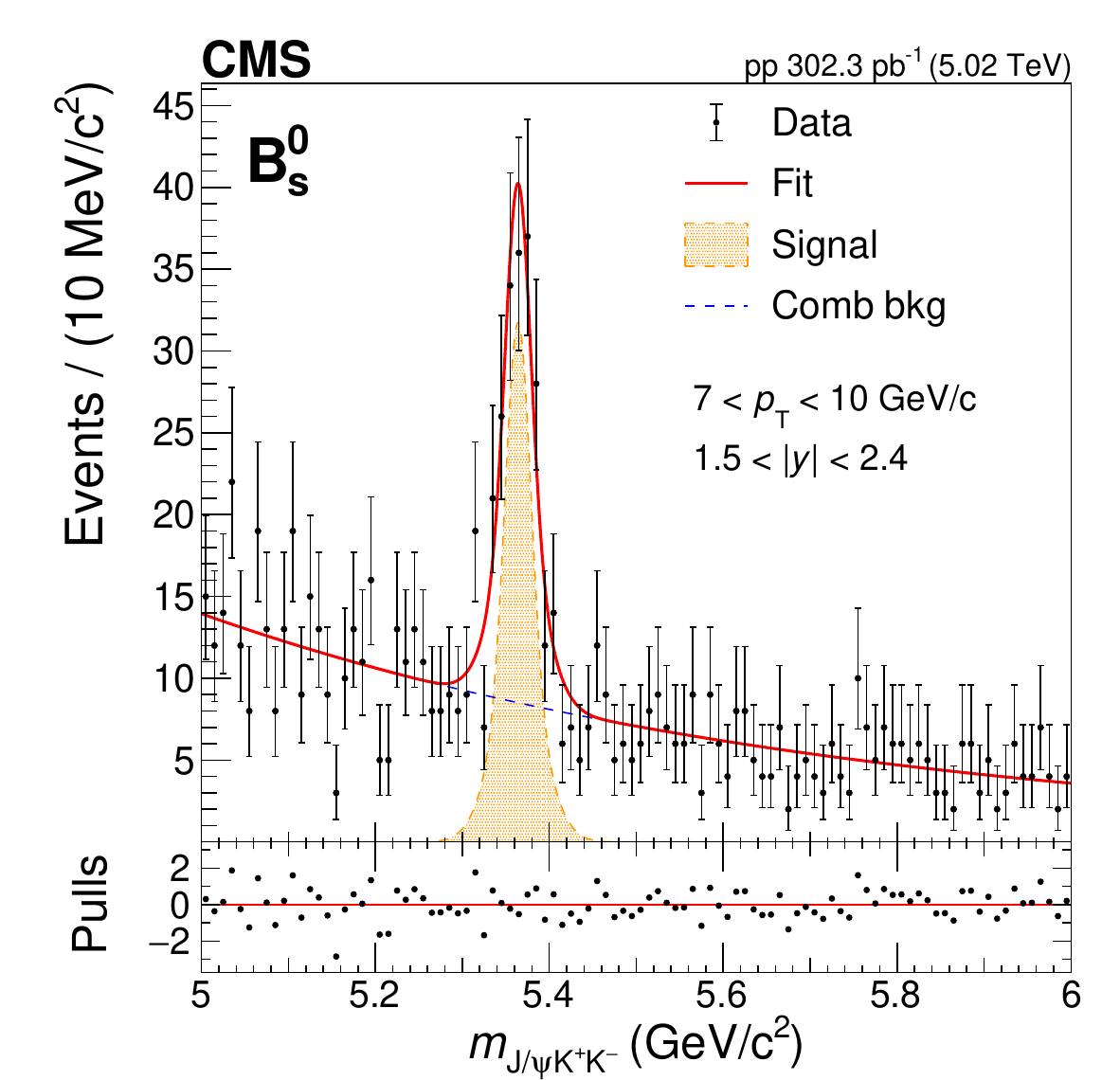}
\includegraphics[width=.49\textwidth]{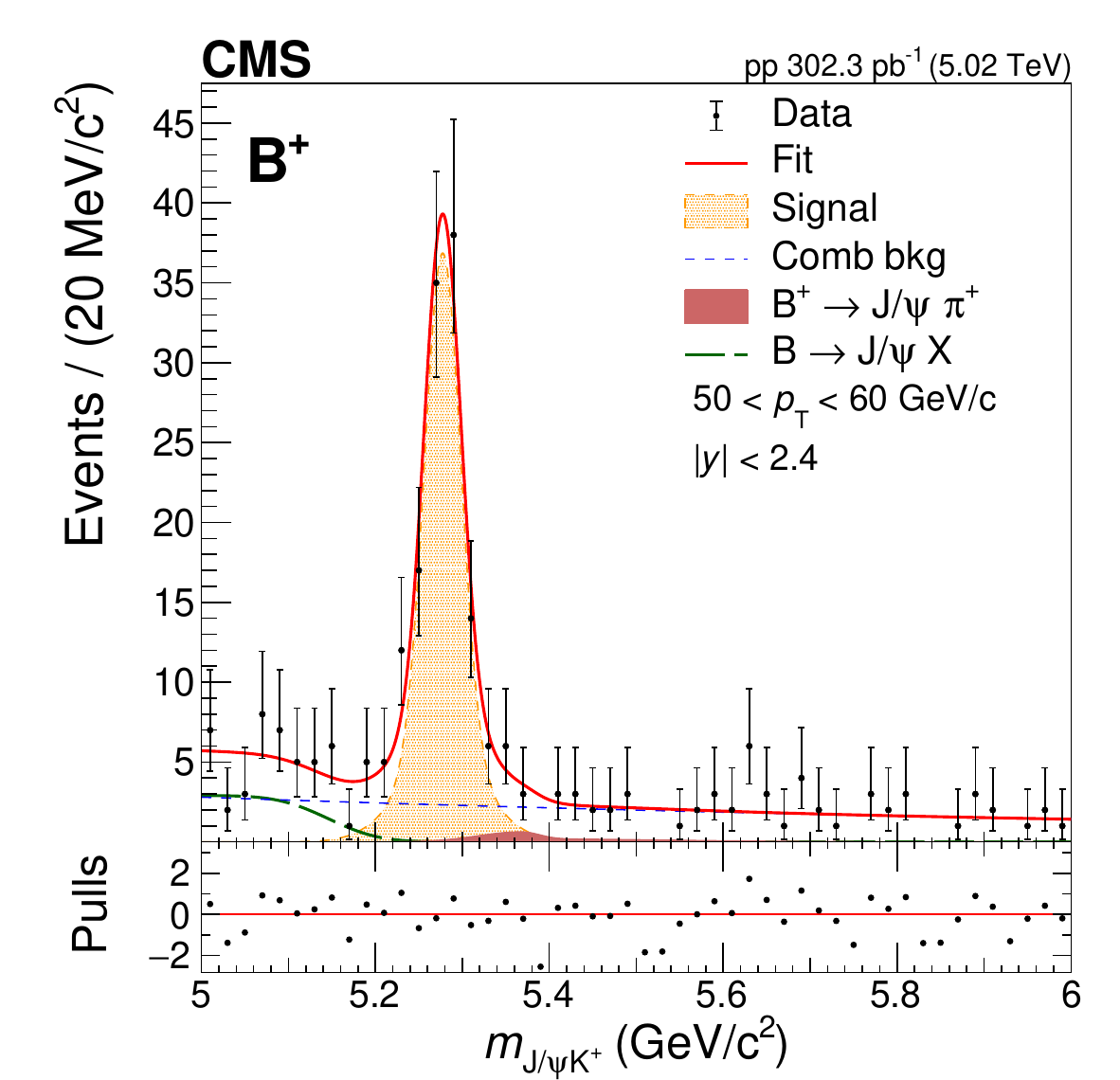}
\includegraphics[width=.49\textwidth]{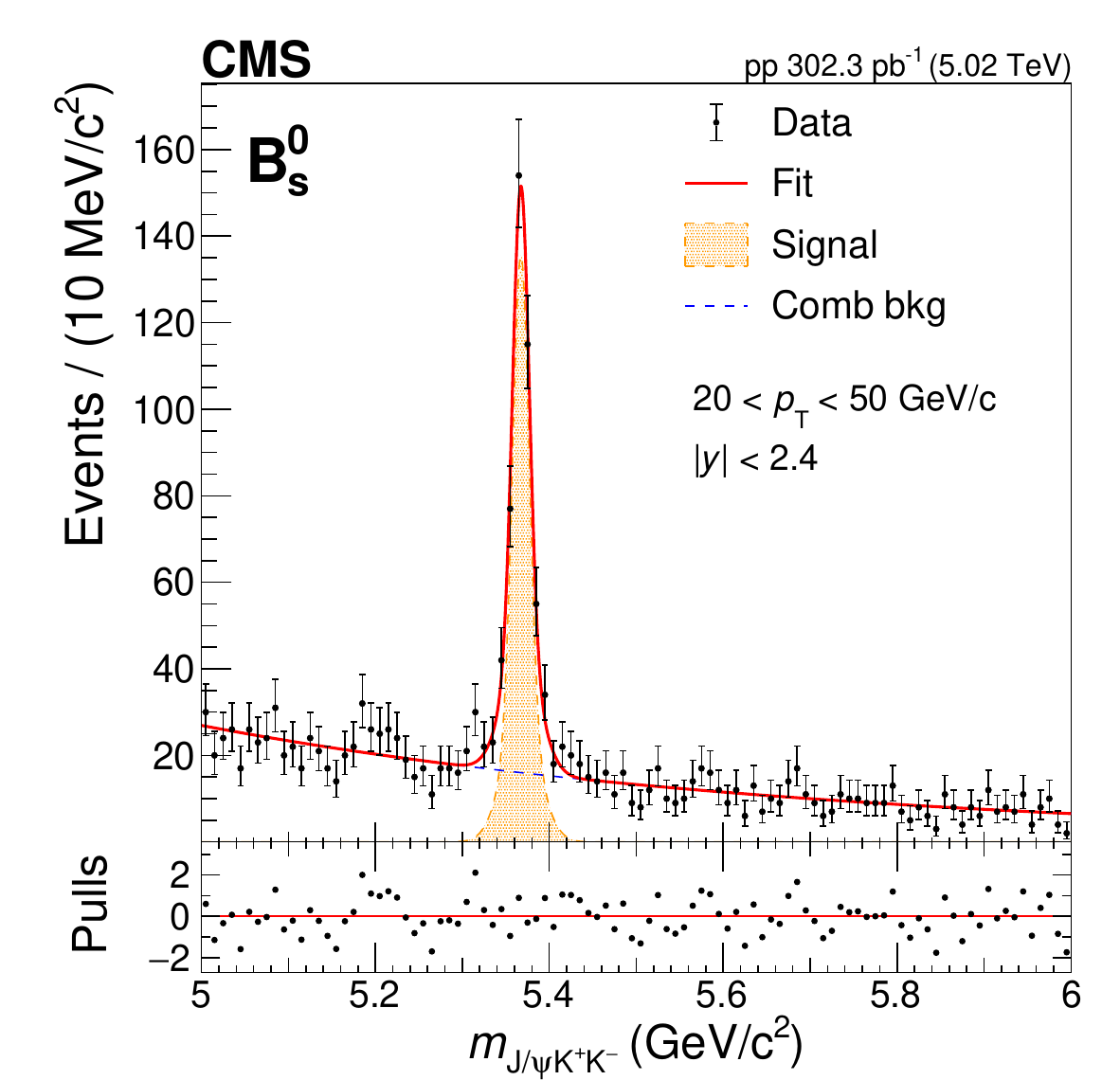}
\caption{Invariant mass distributions of \PBp (\cmsLeft) and \PBzs (\cmsRight) candidates, 
for the lowest (upper) and highest (lower) $\pt$ ranges in \pp
  collisions at $\sqrts=5.02\TeV$. The bottom section in each panel shows the pulls, calculated as the difference between the data points and the fit results, divided by the uncertainty in data.}
\label{fig:massPeaks}
\end{figure}

Additional background sources can arise from possible contamination from other \PB hadron decays, $\PB \to \PJGy\PX$. 
The Cabibbo-suppressed decay $\PBp \to \PJGy\,\Pgpp$, where the pion is incorrectly
assigned the kaon mass, results in a small peaking structure in the signal region. It is modeled by the sum of two split normal distribution functions, with both the shape and the normalization (relative to the signal) being fixed in the fit to the data.
Partially reconstructed decays lead to an accumulation of events in the low mass sideband that is also modeled from simulation and described by an error function. 
For the \PBzs meson, such misreconstructed and partially reconstructed background contributions are found to be negligible. 
Examples of the signal extraction from \PBp and \PBzs invariant-mass distributions, corresponding to the lowest and highest \pt ranges, 
are shown in Fig.~\ref{fig:massPeaks}. 

{\tolerance=800
The differential cross section for \PB
meson production in \pp collisions is calculated in each \pt interval according to
\begin{equation}
 \frac{{\rd}\sigma_{\pp}}{{\rd}\pt} =
 \frac{1}{2} \frac{N_{\text{obs}}(\pt)}{\mathcal{B} \, \mathcal{L}} \frac{1}{\Delta\pt} 
 \Big\langle \frac{
 1}{\alpha(\pt, y) \, \epsilon(\pt, y)} \Big\rangle,
 \label{eq:crosssectionDpPb}
\end{equation}
where 
$N_{\text{obs}}$ 
is the raw signal yield extracted in each \pt interval of width $\Delta \pt$, $\mathcal{B}$ is the branching fraction of the corresponding decay chain obtained from Ref.~\cite{10.1093/ptep/ptac097}, and $\mathcal{L}$ is the integrated luminosity. 
The factor $\frac{1}{2}$  is to quote the production cross sections for particles, whereas the raw yields include both particles and antiparticles.
The acceptance and efficiency  factor, $\langle 1/(\alpha(\pt, y)\epsilon(\pt, y)) \rangle$, is obtained employing
two-dimensional (2D),
fine-grained maps of acceptance $\alpha(\pt, y)$ and efficiency $\epsilon(\pt, y)$.
The acceptance corresponds to the fraction of generated signal events passing the muon and
kaon selection thresholds. The efficiency is determined as the fraction
of accepted signal events that pass the complete analysis selection criteria,
including losses resulting from inefficiencies in the event trigger.
These maps are determined from MC-simulated samples of \PB meson signal events generated within the fiducial region in the analysis. 
The maps are used to determine the $1/(\alpha\epsilon)$ value for each \PB candidate in
data, based on the kinematics $(\pt, y)$ of the candidates. The corresponding average
$\langle 1/(\alpha\epsilon)\rangle$ was obtained for the candidates within a
$\pm 80\MeVcc$ window centered on the \PB meson's world-average mass.
Combinatorial background can influence the \pt and \(y\) dependent correction factor by distorting the average $1/(\alpha\epsilon)$ value from that which would be obtained if this background could be excluded. The impact of combinatorial background is studied and included as a systematic uncertainty.
The efficiency maps derived from the simulation are corrected by data-to-simulation scale factors for the muon reconstruction and trigger efficiencies, obtained by applying the "tag-and-probe" method~\cite{Khachatryan:2010zg} using the \PJGy resonance.
\par}

\section{Systematic uncertainties}
The measured \PBp and \PBzs cross sections in \pp collisions are affected by several systematic uncertainties arising from the signal extraction, the efficiency and acceptance corrections, the associated branching fractions, and the integrated luminosity measurement.

The uncertainty in the raw yield extraction is evaluated by considering the following fitting model variations : 
(i) using a sum of three Gaussian functions with a common mean, 
(ii) using a sum of a Gaussian function and a Crystal Ball function~\cite{slacthe},
(iii) fixing the double-Gaussian mean to its MC simulation value
for the signal modeling instead of letting it be a free parameter,
(iv) using low-order polynomials for describing the combinatorial background, and
(v) changing the fitting range from 5--6\GeVcc to 5.19--6\GeVcc to minimize the $\PBp \to \PJGy\PX$ component
in the background modeling.
For each \pt interval, we take the maximum of the percent difference between the varied and the nominal yield among (i)--(iii) as the signal uncertainty
and (iv)--(v) as the background uncertainty.
The signal and background uncertainties are added in quadrature as the systematic uncertainty in raw yield extraction.

The efficiency and acceptance determinations are affected by potential disagreements between data and simulation for the \PB meson selection efficiency, the track reconstruction efficiency, and the efficiency of the muon trigger, reconstruction, and identification. The systematic uncertainties associated with these effects are estimated as done for the \PbPb analysis results~\cite{CMS:2021mzx}.
To account for the potential discrepancy between data and simulation, the signal MC simulation is validated by inspecting the distributions of the discriminating variables listed in Section~\ref{sec:selection}.
For each \pt range, the signal distributions are extracted from the data employing the $\sPlot$ method~\cite{splot}, using the mass of the \PB meson candidates as a discriminating variable, and are also cross-checked with a simple sideband subtraction method. In particular, the $\sPlot$-derived signal distributions for the BDT score are retrieved from the data, and the corresponding data-to-simulation ratios computed. These ratios are, in turn, used to re-weight the MC simulation, and the resulting deviation in the \meanacceffinv factors are assigned as systematic uncertainties.

The systematic uncertainty associated with the combinatorial background and its influence on the 2D-map-based correction factor is studied by varying the \PB meson candidate mass window range by a factor of two. The resulting change in the correction factor is quoted as
a 
systematic uncertainty, which varies from 0.2\%--2.8\% for \PBp and from 0.3\%--2.3\% for \PBzs mesons.
The MC simulation is weighted by the FONLL-to-\PYTHIA ratios of \pt distributions, and the difference in the \meanacceffinv factors are assigned as systematic uncertainties to estimate the potential impact of the assumed \PB meson \pt shape in MC. The uncertainty arising from the \pt shape is negligible.
The uncertainty in the efficiency of the muon trigger, reconstruction, and identification is evaluated using the tag-and-probe method~\cite{Khachatryan:2010zg}. The systematic uncertainty in the muon efficiency has an insignificant effect on both the \PBp and \PBzs meson results.
The derived data-to-MC scale factors are employed to determine the nominal $(\acceff)$ 2D maps, while the difference in the \meanacceffinv factors between the scaled and original values are taken as the systematic uncertainty.
The difference in the track reconstruction efficiency in data and simulation is estimated
by comparing 3-prong and 5-prong \PDst decays, \dstardecay\, and \dstardecayfive~\cite{CMS-DP-2018-050}.
 This difference results in a 2.4\% uncertainty in the efficiency determination for each hadronic track, independent of \pt.
 We consider this uncertainty to be uncorrelated between tracks.
In addition, the selections on track quality are tightened and loosened. These selections
include the relative \pt uncertainty, the normalized $\chi^{2}$ by the number of degrees
of freedom, and the sum of the numbers of pixel and strip layer hits. The maximum
resulting deviation of cross sections from the nominal values is quoted as a systematic uncertainty. The uncertainties associated with the branching fraction $\mathcal{B}$ of the decay chains are obtained from Ref.~\cite{10.1093/ptep/ptac097}. 

The total systematic uncertainties in the cross sections are computed as the sum in quadrature of the different contributions mentioned above, which are summarized in Table~\ref{tab:SystSumCombined} for the \PBp and \PBzs cross sections.
The systematic uncertainties resulting from
the fitting model variation,
the discrepancy between the MC simulation and the experimental results for the discriminating variables,
the background contamination of the efficiency,
the muon efficiency,
the track selection,
and the integrated luminosity for \pp and \PbPb are taken as
independent and added in quadrature as the systematic uncertainty in the nuclear
modification factors. On the other hand,
for the \RAA calculations, the branching fractions are common to the \pp and \PbPb systems
and, consequently, do not impact the results. The track reconstruction efficiencies used
for the two systems also have common elements, resulting in partial cancellation of the corresponding uncertainties when calculating the \RAA values.
The global uncertainties in \RAA consist of the uncertainties in \TAA, the number of minimum bias events, and the integrated luminosity in \pp collisions.

\begin{table}[ht!]
\centering
\topcaption{Summary of systematic uncertainties (in \%) for the \PBp (upper table) and \PBzs
  (lower one) meson cross section as a
  function of \pt. The systematic uncertainties in luminosity and branching fractions are global uncertainties that are applied equally to all \pt ranges.
}
\label{tab:SystSumCombined}
  \begin{tabular}{ c c c  c  c  c  c c c}
    \hline
    & \multicolumn{7}{c}{\PBp \pt (\GeVcns)} \\
    Source & 5--7 & 7--10 & 10--15 & 15--20 & 20--30 & 30--50 & 50--60  & 20--50\\
    \hline
    Fitting model variation & 2.1 & 1.4 & 3.2 & 1.1 & 0.69 & 1.8 & 2.4 & 0.57\\[0pt]
    Data-to-simulation discrepancy & 4.7 & 7.2 & 7.2 & 0.98 & 0.87 & 0.92 & 0.83 & 0.84\\[0pt]
    Bkg. contamination: efficiency & 1.5 & 2.8 & 0.84 & 0.41 & 0.46 & 0.18 & 1.1 & 0.41\\[0pt]
    Muon efficiency & 0.47 & 0.45 & 0.37 & 0.36 & 0.43 & 0.64 & 0.64 & 0.47\\[0pt]
    Hadron tracking efficiency & 2.4 & 2.4 & 2.4 & 2.4 & 2.4 & 2.4 & 2.4 & 2.4\\[0pt]
    Track selection & 1.8 & 0.31 & 0.43 & 0.37 & 0.27 & 0.052 & 1.6 & 0.24\\[0pt]
    Sum & 6.2 & 8.3 & 8.3 & 2.9 & 2.7 & 3.2 & 4.1 & 2.7\\[0pt]

    \hline
      Integrated luminosity  & 
      \multicolumn{7}{c}{1.9}\\
    Branching fractions & 
          \multicolumn{7}{c}{2.9}\\
   Sum (global systematic unc.) & 
      \multicolumn{7}{c}{3.5}\\
    \hline
\end{tabular}

\vspace{1em}

  \begin{tabular}{ c c c  c  c }
    \hline
& \multicolumn{4}{c}{\PBzs \pt (\GeVcns)} \\
Source &  7--10 & 10--15 & 15--20 & 20--50  \\
    \hline
    Fitting model variation & 3.6 & 2.0 & 2.9 & 3.2\\[0pt]
    Data-to-simulation discrepancy & 3.7 & 1.9 & 1.7 & 1.5\\[0pt]
    Bkg. contamination: efficiency  & 1.1 & 2.3 & 0.28 & 0.38\\[0pt]
    Muon efficiency & 0.46 & 0.38 & 0.35 & 0.45\\[0pt]
    Hadron tracking efficiency & 4.8 & 4.8 & 4.8 & 4.8\\[0pt]
    Track selection & 0.65 & 0.2 & 2.7 & 0.78\\[0pt]
    Sum & 7.2 & 6.0 & 6.5 & 6.0\\[0pt]
    \hline
      Integrated luminosity  & 
    \multicolumn{4}{c}{1.9} \\
    Branching fractions & 
    \multicolumn{4}{c}{7.5} \\
   Sum (global systematic unc.) & 
    \multicolumn{4}{c}{7.7} \\
    \hline
\end{tabular}
\end{table}

\section{Results}

\begin{figure}[!ht]
\centering
\includegraphics[width=.45\textwidth]{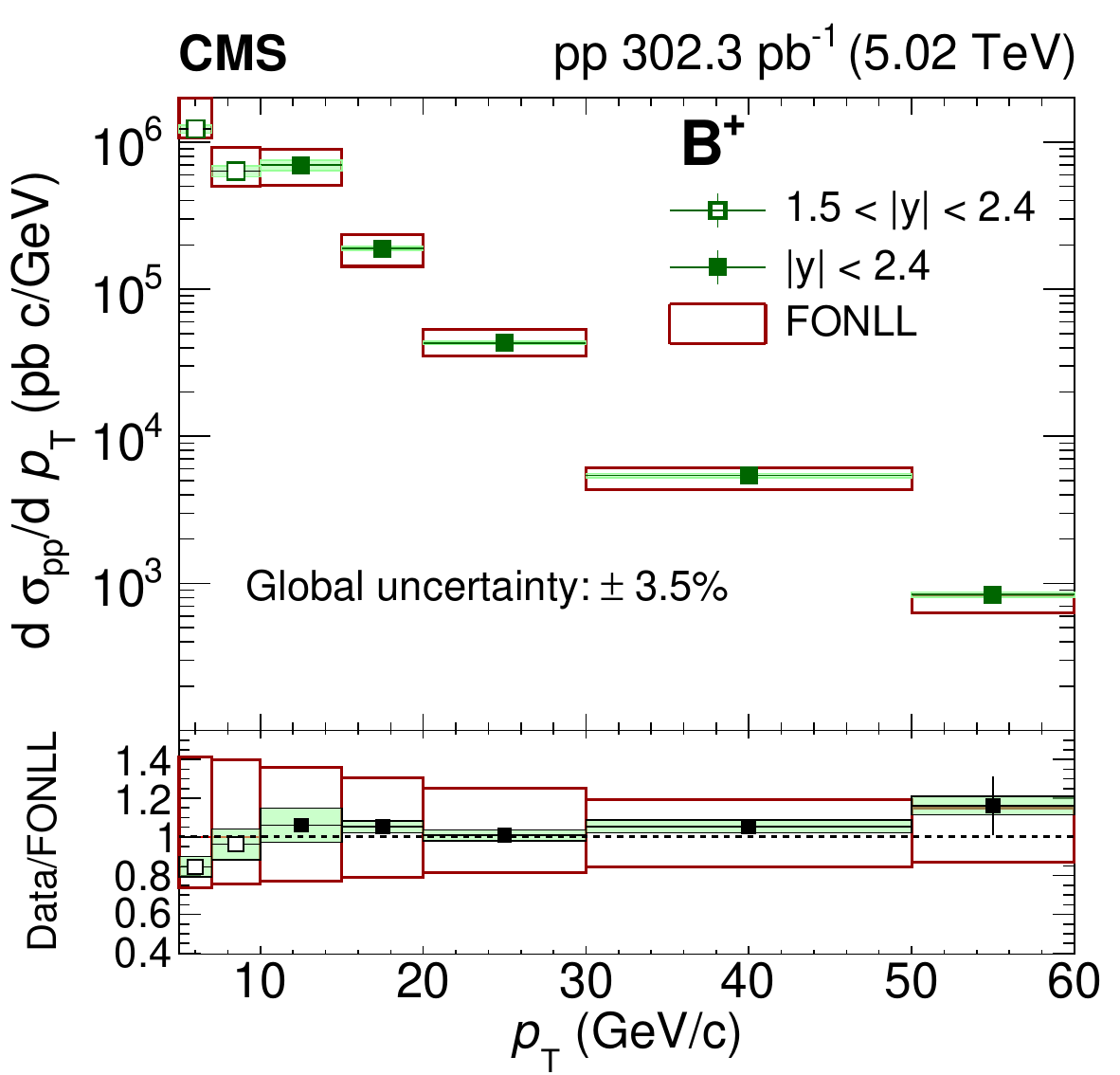}
\includegraphics[width=.45\textwidth]{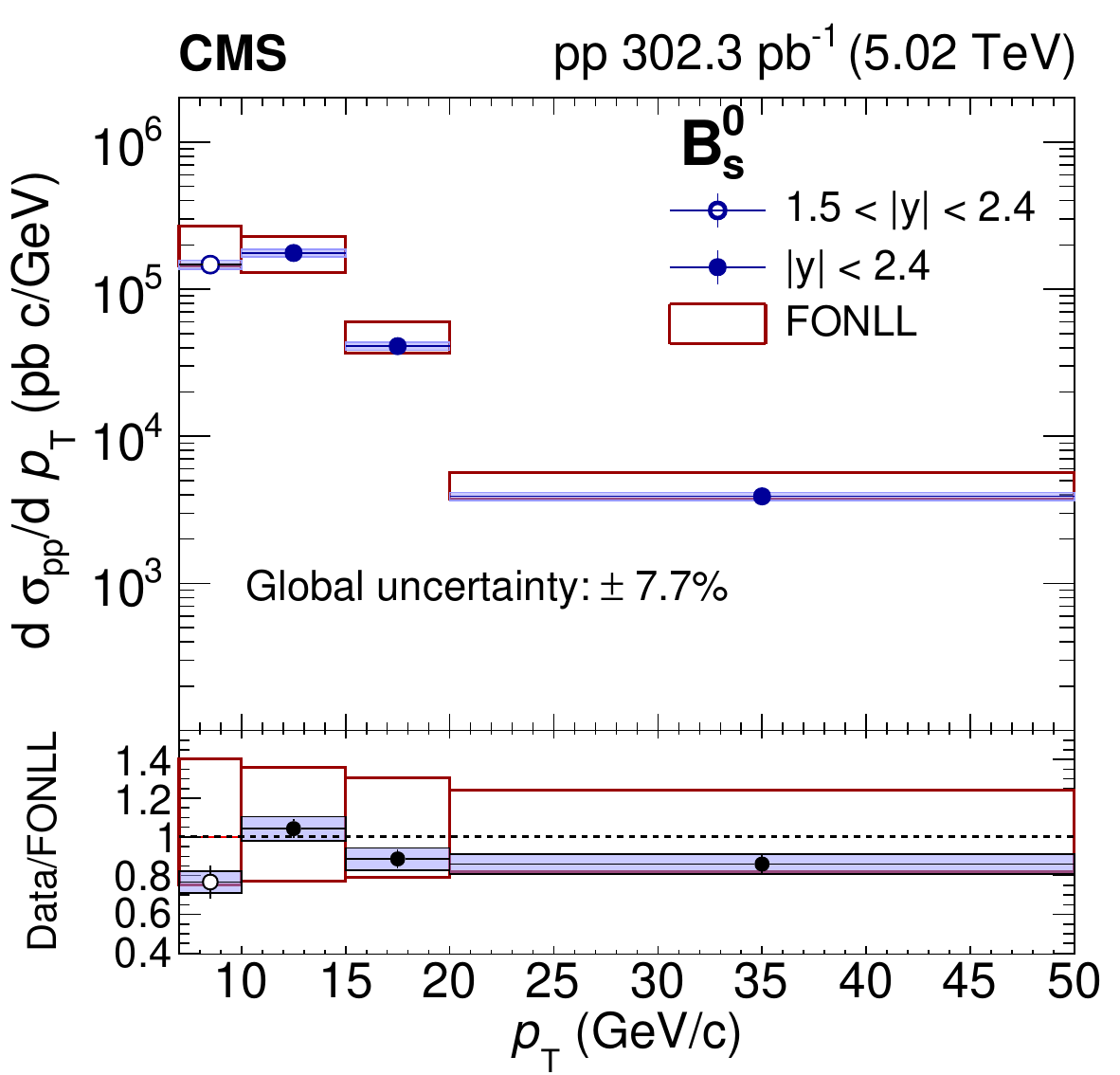} 
\caption{The \pt-differential cross sections of \PBp (left) and \PBzs (right) mesons in \pp
  collisions at $\sqrts=5.02\TeV$. The vertical bars (shaded boxes) correspond to
  statistical (systematic) uncertainties.
  The horizontal bars reflect the bin widths.
  The global systematic uncertainty includes uncertainties in the integrated luminosity and the external branching fractions. The comparison with the calculation from
  FONLL (open boxes) is also shown.
  The lower panels show that the measured cross sections deviate from the FONLL calculations by less than 20\%, and the two are consistent within uncertainty.}
\label{fig:xsections}
\end{figure}

The \pt-differential production cross sections of \PBp and \PBzs mesons are presented in
Fig.~\ref{fig:xsections}. The results are compared to the cross sections obtained from
FONLL calculations~\cite{FONLLcharmbottomPP3}, which are obtained by scaling the FONLL
total \PQb quark
production~\cite{FONLLcharmbottomPP1,FONLLcharmbottomPP2,FONLLcharmbottomPP3} by the
production fractions at high energy of 40.1\% for \PBp mesons~\cite{Beringer:1900zz} and
10.3\% for \PBzs mesons~\cite{HFLAV:2016hnz}.
The calculated FONLL reference spectra are consistent with the \pp spectra measured for both the \PBp and \PBzs mesons within the quoted uncertainties.
With smaller experimental uncertainties, these measurements are useful for constraining the \PQb-quark fragmentation functions.

\begin{figure}[!h]
\centering
\includegraphics[width=.46\textwidth]{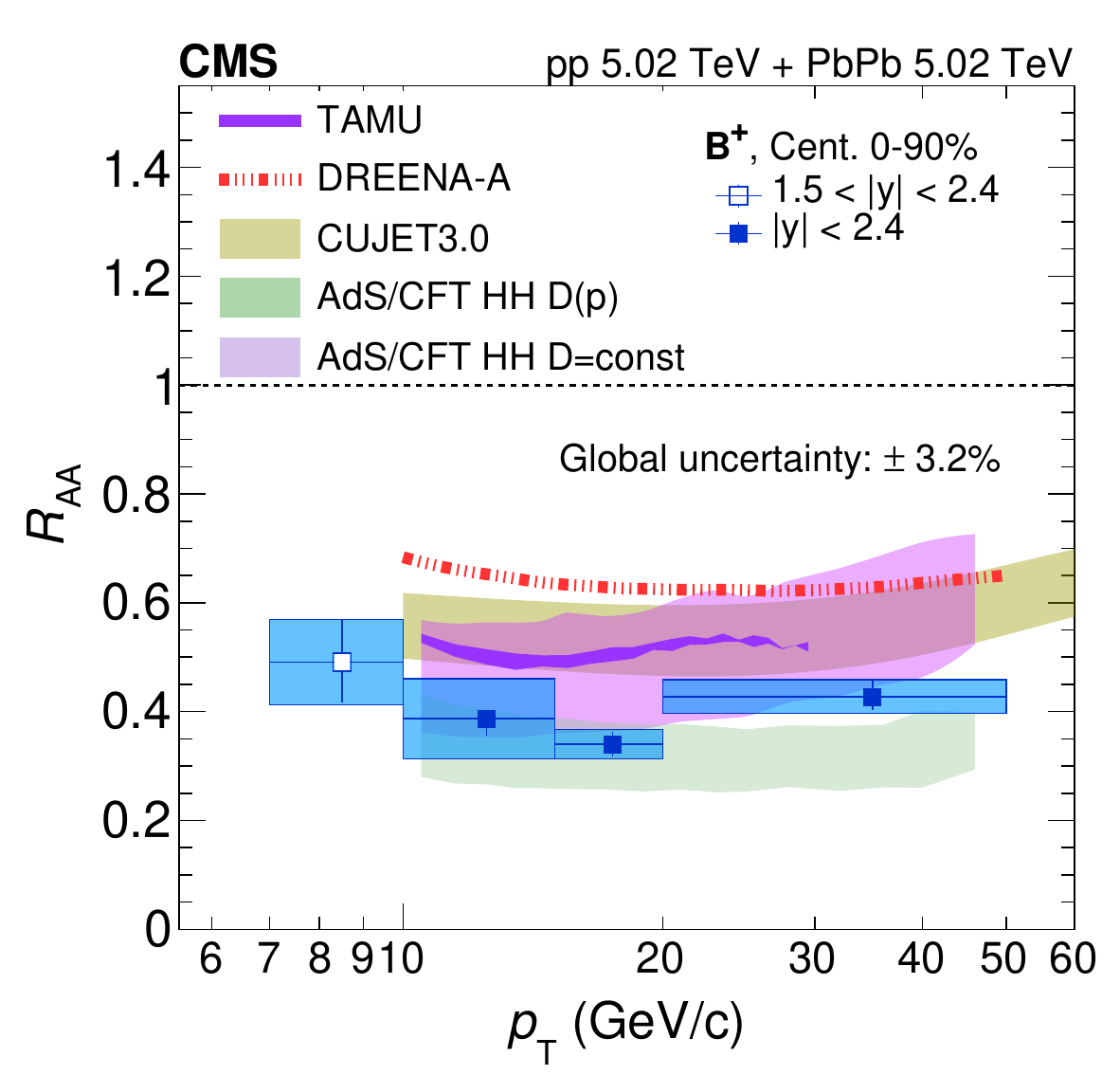}
\includegraphics[width=.46\textwidth]{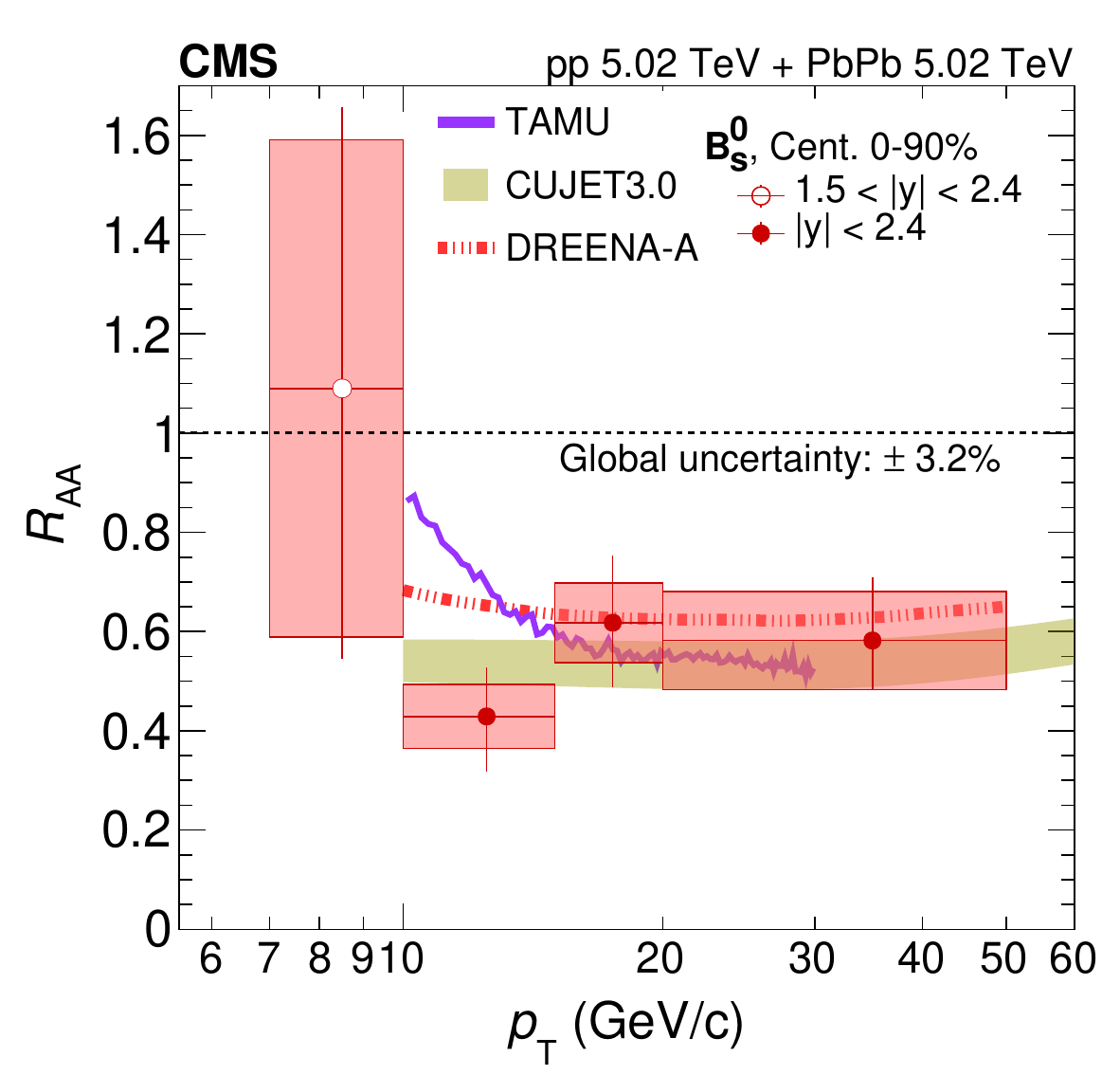}
\caption{The \pt dependence of the nuclear modification factor of \PBp (left) and \PBzs (right) mesons using \PbPb and \pp collisions at $\sqrtsNN=5.02\TeV$. The vertical bars (boxes) correspond to statistical (systematic) uncertainties. The horizontal bars reflect the bin widths and not uncertainties.The global systematic uncertainty includes the uncertainties in the integrated luminosity, \TAA, and the number of minimum bias events~\cite{CMS:2021mzx}. For the \PBp meson, four theoretical calculations are also shown for comparison:
  TAMU~\cite{He:2011qa, He:2014cla}, DREENA-A~\cite{Zigic:2022xks, Zigic:2021rku}, CUJET3.0~\cite{Xu:2015bbz, Xu:2014tda, Xu:2014ica}, and AdS/CFT HH~\cite{Horowitz:2015dta,  AdscftHH} with the diffusion coefficient either dependent or independent of the quark momentum.
  For the \PBzs meson, three theoretical calculations are shown for comparison:
  TAMU~\cite{He:2011qa, He:2014cla}, DREENA-A~\cite{Zigic:2022xks, Zigic:2021rku}, and CUJET3.0~\cite{Xu:2015bbz, Xu:2014tda, Xu:2014ica}.
  The line width of the theoretical calculation from Refs.~\cite{He:2011qa, He:2014cla} indicates the uncertainty in the total \PQb quark coalescence probability (no shadowing is applied). The rapidity range is $\abs{y} < 2.4$ for all the theoretical predictions and, therefore, these predictions can only be compared to the measurements in the range of $\pt > 10\GeVc$.
}
\label{fig:RAA}
\end{figure}

The nuclear modification factors of \PBp and \PBzs, determined according to Eq.~\eqref{eq:RAA}, are shown in Fig.~\ref{fig:RAA} as a functions of \pt.
For the \PBp meson a strong suppression is observed in \PbPb collisions (with \RAA values significantly below unity).
In the range $7<\pt<50\GeVc$, \RAA varies from 0.3 to 0.6.
For the \PBzs meson, the central value is found to be larger than 1 (\ie, a production enhancement) in the range $7<\pt<10\GeVc$, although the large uncertainty suggests that it is also consistent with the \RAA values at higher \pt.
In the range $10<\pt<50\GeVc$ the \RAA values are also smaller than unity, but larger than those found for the \PBp meson.
This observation, together with the trend of \PBzs production enhancement observed over higher \pt ranges, suggests that recombination may play an important role in the production of \PBzs mesons, especially
at low \pt, although reduced uncertainties for the \PbPb results are needed to establish this effect.

The \pt dependence of the \PBp \RAA values are compared to the predictions of three types of models:
(a) two perturbative QCD-based models that include both collisional and radiative energy
loss (DREENA-A~\cite{Zigic:2022xks, Zigic:2021rku} and CUJET3.0~\cite{Xu:2015bbz, Xu:2014tda, Xu:2014ica}); (b) a transport model based on a Langevin equation featuring collisional energy loss and heavy quark diffusion in the medium (TAMU~\cite{He:2011qa, He:2014cla}), with the spatial diffusion coefficient $D_{s}$ characterizing the long-wavelength limit of heavy-flavor transport~\cite{Dong:2019byy}, and (c) a model based on the anti-de-Sitter/conformal field theory correspondence, that includes thermal fluctuations in the energy loss for heavy quarks in a strongly coupled plasma (AdS/CFT HH)~\cite{Horowitz:2015dta, AdscftHH}.
The AdS/CFT HH calculation is provided for two settings of the diffusion coefficient \textit{D} of the heavy-quark propagation through the medium, either dependent on or independent of the quark momentum. 
Apart from the interactions between heavy quarks and the medium, the set of the
(nuclear) parton distribution functions used for the initial heavy-quark \pt distributions
are also different between these models. Furthermore, the difference includes the modeling of the \PbPb medium (hydrodynamically~\cite{Xu:2015bbz, He:2014cla} or via a Glauber model~\cite{Zigic:2022xks, Zigic:2021rku}) and the energy loss sources (partonic
only~\cite{Xu:2015bbz, He:2014cla} or also hadronic~\cite{He:2014cla}).

The TAMU model, which does not include radiative energy loss,
overpredicts the observed \PBp \RAA values over the available $\pt > 10\GeVc$ range, where the radiative energy loss is expected to play a more significant role.
The perturbative QCD-based models CUJET3.0 and DREENA-A both predict slightly weaker
suppression than TAMU, despite the inclusion of radiative energy loss,
with DREENA-A predicting the least suppression in the $10 < \pt < 15\GeVc$ range.
The AdS/CFT HH calculation with the diffusion coefficient dependent on the quark momentum has the broadest agreement with the data,
although it slightly underestimates the \RAA value in the highest \pt range.
The \RAA values are compatible with the lower-end values predicted by AdS/CFT HH calculation with the diffusion coefficient independent of the quark momentum.
The experimental \RAA value for the highest \pt range falls between the two AdS/CFT HH model predictions.

Given that the experimental uncertainties are smaller than the uncertainties of the theoretical calculations from CUJET3.0 and AdS/CFT HH,
the current results could be used to optimize parameter settings in such models (\eg, the parton-medium coupling parameters in the AdS/CFT HH  model) and constrain the relevance of collisional and radiative processes in the \PQb quark energy loss~\cite{Rohrmoser:2016yct,CAO2014569}. 
The \PBzs \RAA result is compared to DREENA-A, TAMU, and CUJET3.0. The difference between TAMU and CUJET3.0 below $\pt=15\GeVc$ reflects the contribution from recombination processes, which are included in the
TAMU model, but are not present in the CUJET3.0 model. Given the current statistical and systematic uncertainties, these three theoretical predictions are roughly compatible with the measurement. 

\begin{figure}[!h]
  \centering
  \includegraphics[width=.46\textwidth]{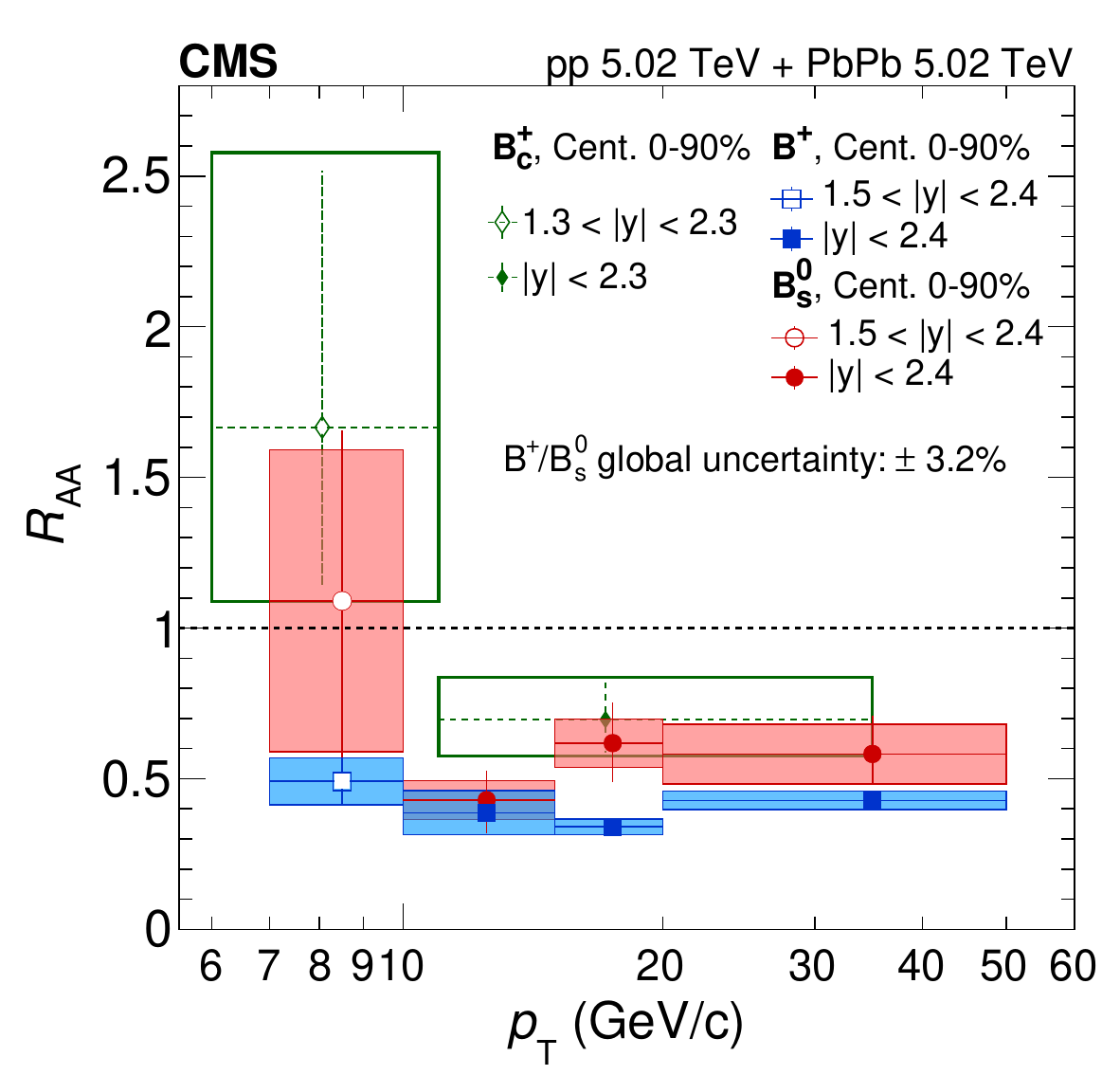}
  \includegraphics[width=.46\textwidth]{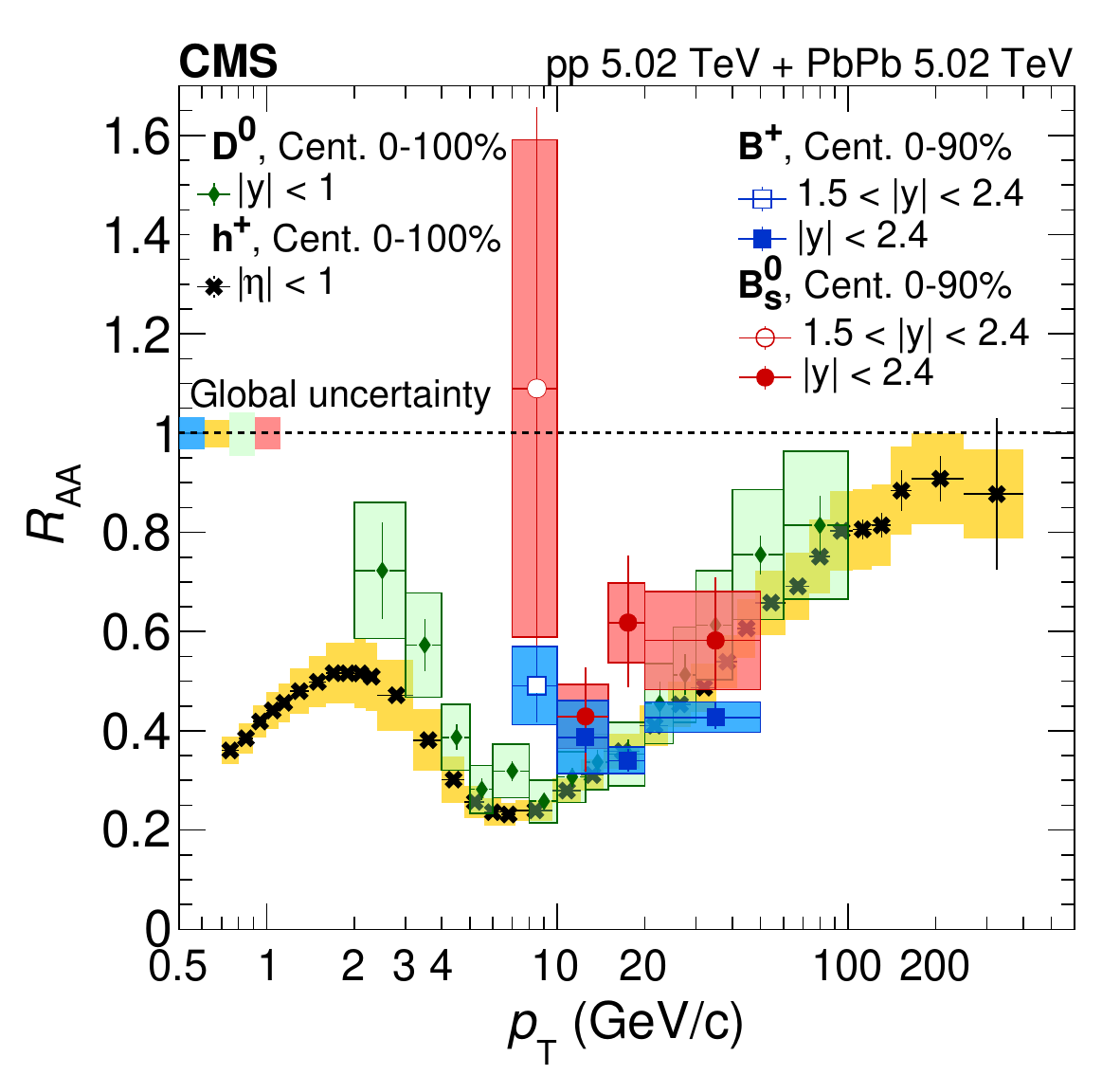}
  \caption{Comparison of the nuclear modification factor as a function of \pt of \PBp and \PBzs mesons with \PBc mesons~\cite{CMS:2022sxl} (left) as well as with \PDz mesons~\cite{CMS:2018474} and inclusive charged hadrons~\cite{CMS:2016xef} (right) in \PbPb collisions at $\sqrtsNN=5.02\TeV$.
  The vertical bars (boxes) correspond to statistical (systematic) uncertainties.
  The horizontal bars reflect the bin widths.
    The global systematic uncertainties,
    written in text on the left plot and
    depicted in shaded boxes around $\RAA=1$ on the right plot, comprise the
    uncertainties in \TAA, the number of minimum-bias events, and the luminosity in \pp collisions.
    The dashed lines and the open boxes in the \PBc data (left plot) represent bin-to-bin-uncorrelated and total uncertainties, respectively.
  }
  \label{fig:RAA_light}
\end{figure}

The \PBp and \PBzs \RAA are compared in Fig.~\ref{fig:RAA_light} to the CMS measurements
of the \RAA of  \PBc mesons~\cite{CMS:2022sxl}, \PDz mesons~\cite{CMS:2018474}, and
charged particles~\cite{CMS:2016xef} performed
at the same energy and in either the same (0--90\%) or a similar (0--100\%) centrality range. For the heavy \PBc mesons, the rapidity is similarly limited to the forward region at low \pt, and for the lighter charged particles and \PDz mesons the measurement is performed at mid-rapidity.
The \RAA value for \PBc mesons in the $7 < \pt < 50\GeVc$ range is higher than that for \PBp mesons, with \PBzs values falling in between.
For \PBp mesons, the \RAA values are consistent with those of charged particles and \PDz
mesons for $\pt > 10\GeVc$ . At lower \pt, a reduced level of suppression is observed,
which is in line with expectations based on the quark mass dependence of parton energy
loss~\cite{Apolinario:2022vzg, Dokshitzer:1991fd}.

\section{Summary}
In summary, the differential cross sections of \PBp and \PBzs mesons as functions of
transverse momentum (\pt) in proton-proton (\pp) collisions at a center-of-mass energy of
$5.02\TeV$ have been measured with the CMS detector at the LHC. The mesons were
reconstructed via the exclusive decay channels \Bplusdecayshort and \Bzerosdecayshort in
the \pt intervals of 5--60 and 7--50\GeVc for \PBp and \PBzs, respectively. The measured \pt-differential cross sections of \PBp and \PBzs mesons in \pp collisions are well-described by fixed-order plus next-to-leading logarithm calculations.

The nuclear modification factors (\RAA) are obtained from these differential cross sections and the corresponding measurements in lead-lead collisions.
The \RAA of \PBzs and \PBp mesons are significantly lower than unity for $\pt > 10 \GeVc$, while at lower \pt the \PBzs meson results hint at having larger \RAA values than found for \PBp mesons.
The \RAA data for \PBp and \PBzs mesons are compared to theoretical calculations, providing constraints for the mechanisms of bottom quark energy loss and hadronization in the quark-gluon plasma. The results are also compared to experimental \RAA values for charm and light-flavor hadrons.
The \PBp meson \RAA values are compatible with those found for generic charged hadrons and \PDz mesons in the \pt range from 10 to 30\GeVc,
while the \RAA values from 7 to 10\GeVc are consistent with expectations based on the quark mass dependence of parton energy loss.

\begin{acknowledgments}
We congratulate our colleagues in the CERN accelerator departments for the excellent performance of the LHC and thank the technical and administrative staffs at CERN and at other CMS institutes for their contributions to the success of the CMS effort. In addition, we gratefully acknowledge the computing centers and personnel of the Worldwide LHC Computing Grid and other centers for delivering so effectively the computing infrastructure essential to our analyses. Finally, we acknowledge the enduring support for the construction and operation of the LHC, the CMS detector, and the supporting computing infrastructure provided by the following funding agencies: SC (Armenia), BMBWF and FWF (Austria); FNRS and FWO (Belgium); CNPq, CAPES, FAPERJ, FAPERGS, and FAPESP (Brazil); MES and BNSF (Bulgaria); CERN; CAS, MoST, and NSFC (China); MINCIENCIAS (Colombia); MSES and CSF (Croatia); RIF (Cyprus); SENESCYT (Ecuador); ERC PRG, RVTT3 and MoER TK202 (Estonia); Academy of Finland, MEC, and HIP (Finland); CEA and CNRS/IN2P3 (France); SRNSF (Georgia); BMBF, DFG, and HGF (Germany); GSRI (Greece); NKFIH (Hungary); DAE and DST (India); IPM (Iran); SFI (Ireland); INFN (Italy); MSIP and NRF (Republic of Korea); MES (Latvia); LMTLT (Lithuania); MOE and UM (Malaysia); BUAP, CINVESTAV, CONACYT, LNS, SEP, and UASLP-FAI (Mexico); MOS (Montenegro); MBIE (New Zealand); PAEC (Pakistan); MES and NSC (Poland); FCT (Portugal); MESTD (Serbia); MCIN/AEI and PCTI (Spain); MOSTR (Sri Lanka); Swiss Funding Agencies (Switzerland); MST (Taipei); MHESI and NSTDA (Thailand); TUBITAK and TENMAK (Turkey); NASU (Ukraine); STFC (United Kingdom); DOE and NSF (USA).

\hyphenation{Rachada-pisek} Individuals have received support from the Marie-Curie program and the European Research Council and Horizon 2020 Grant, contract Nos.\ 675440, 724704, 752730, 758316, 765710, 824093, 101115353, 101002207, and COST Action CA16108 (European Union); the Leventis Foundation; the Alfred P.\ Sloan Foundation; the Alexander von Humboldt Foundation; the Science Committee, project no. 22rl-037 (Armenia); the Belgian Federal Science Policy Office; the Fonds pour la Formation \`a la Recherche dans l'Industrie et dans l'Agriculture (FRIA-Belgium); the F.R.S.-FNRS and FWO (Belgium) under the ``Excellence of Science -- EOS" -- be.h project n.\ 30820817; the Beijing Municipal Science \& Technology Commission, No. Z191100007219010 and Fundamental Research Funds for the Central Universities (China); the Ministry of Education, Youth and Sports (MEYS) of the Czech Republic; the Shota Rustaveli National Science Foundation, grant FR-22-985 (Georgia); the Deutsche Forschungsgemeinschaft (DFG), among others, under Germany's Excellence Strategy -- EXC 2121 ``Quantum Universe" -- 390833306, and under project number 400140256 - GRK2497; the Hellenic Foundation for Research and Innovation (HFRI), Project Number 2288 (Greece); the Hungarian Academy of Sciences, the New National Excellence Program - \'UNKP, the NKFIH research grants K 131991, K 133046, K 138136, K 143460, K 143477, K 146913, K 146914, K 147048, 2020-2.2.1-ED-2021-00181, and TKP2021-NKTA-64 (Hungary); the Council of Science and Industrial Research, India; ICSC -- National Research Center for High Performance Computing, Big Data and Quantum Computing and FAIR -- Future Artificial Intelligence Research, funded by the NextGenerationEU program (Italy); the Latvian Council of Science; the Ministry of Education and Science, project no. 2022/WK/14, and the National Science Center, contracts Opus 2021/41/B/ST2/01369 and 2021/43/B/ST2/01552 (Poland); the Funda\c{c}\~ao para a Ci\^encia e a Tecnologia, grant CEECIND/01334/2018 (Portugal); the National Priorities Research Program by Qatar National Research Fund; MCIN/AEI/10.13039/501100011033, ERDF ``a way of making Europe", and the Programa Estatal de Fomento de la Investigaci{\'o}n Cient{\'i}fica y T{\'e}cnica de Excelencia Mar\'{\i}a de Maeztu, grant MDM-2017-0765 and Programa Severo Ochoa del Principado de Asturias (Spain); the Chulalongkorn Academic into Its 2nd Century Project Advancement Project, and the National Science, Research and Innovation Fund via the Program Management Unit for Human Resources \& Institutional Development, Research and Innovation, grant B39G670016 (Thailand); the Kavli Foundation; the Nvidia Corporation; the SuperMicro Corporation; the Welch Foundation, contract C-1845; and the Weston Havens Foundation (USA).
\end{acknowledgments}

\bibliography{auto_generated}
\cleardoublepage \appendix\section{The CMS Collaboration \label{app:collab}}\begin{sloppypar}\hyphenpenalty=5000\widowpenalty=500\clubpenalty=5000
\cmsinstitute{Yerevan Physics Institute, Yerevan, Armenia}
{\tolerance=6000
A.~Hayrapetyan, A.~Tumasyan\cmsAuthorMark{1}\cmsorcid{0009-0000-0684-6742}
\par}
\cmsinstitute{Institut f\"{u}r Hochenergiephysik, Vienna, Austria}
{\tolerance=6000
W.~Adam\cmsorcid{0000-0001-9099-4341}, J.W.~Andrejkovic, T.~Bergauer\cmsorcid{0000-0002-5786-0293}, S.~Chatterjee\cmsorcid{0000-0003-2660-0349}, K.~Damanakis\cmsorcid{0000-0001-5389-2872}, M.~Dragicevic\cmsorcid{0000-0003-1967-6783}, P.S.~Hussain\cmsorcid{0000-0002-4825-5278}, M.~Jeitler\cmsAuthorMark{2}\cmsorcid{0000-0002-5141-9560}, N.~Krammer\cmsorcid{0000-0002-0548-0985}, A.~Li\cmsorcid{0000-0002-4547-116X}, D.~Liko\cmsorcid{0000-0002-3380-473X}, I.~Mikulec\cmsorcid{0000-0003-0385-2746}, J.~Schieck\cmsAuthorMark{2}\cmsorcid{0000-0002-1058-8093}, R.~Sch\"{o}fbeck\cmsorcid{0000-0002-2332-8784}, D.~Schwarz\cmsorcid{0000-0002-3821-7331}, M.~Sonawane\cmsorcid{0000-0003-0510-7010}, S.~Templ\cmsorcid{0000-0003-3137-5692}, W.~Waltenberger\cmsorcid{0000-0002-6215-7228}, C.-E.~Wulz\cmsAuthorMark{2}\cmsorcid{0000-0001-9226-5812}
% ...
\par}
\vskip\cmsinstskip
\dag:~Deceased\\
$^{1}$Also at Yerevan State University, Yerevan, Armenia\\
$^{2}$Also at TU Wien, Vienna, Austria\\
% ...
\end{sloppypar}
\end{document}